\documentstyle[11pt]{article}
\begin{document}

\title{{\bf Large Gauge Ward Identity}}
\author{Ashok Das \\
\\
{\it Department of Physics and Astronomy, University of Rochester,
Rochester, NY 14627}\\
\\
Gerald Dunne\footnote{permanent address: Department of Physics,
University of Connecticut, Storrs, CT 06269} \\
\\
{\it Department of Physics, Technion - Israel Institute of Technology,
Haifa 32000, Israel}\\
\\
J. Frenkel\\
\\
{\it Instituto de F\'{\i}sica, Universidade de S\~{a}o Paulo, S\~{a}o
Paulo, SP 05315-970, Brazil}.}
\date{}
\maketitle

\begin{abstract}

We study the question of the Ward identity for \lq\lq large'' gauge
invariance in $0+1$ dimensional theories. We derive the relevant Ward
identities for a single flavor fermion and a single flavor complex
scalar field interacting with an Abelian gauge field. These identities
are nonlinear. The Ward identity for any other complicated theory can
be derived from these basic sets of identities. However, the structure
of the Ward identity changes since these are nonlinear identities. In
particular, we work out the \lq\lq large'' gauge Ward identity for a
supersymmetric theory involving a single flavor of fermion as well as
a complex scalar field. Contrary to the effective action for the
individual theories, the solution of the Ward identity in the
supersymmetric theory involves an infinity of Fourier component
modes. We comment on which features of this analysis are
likely/unlikely to generalize to the $2+1$ dimensional theory.
\end{abstract}

\vfill\eject

\section{Introduction:}

The question of \lq\lq large'' gauge invariance at finite temperature
has been
a very interesting one for several years now \cite{LH}. It is well known
that
the
Chern-Simons action in an odd dimensional non-Abelian gauge theory is
not invariant under \lq\lq large'' gauge transformations
\cite{DJT}.  Rather, it shifts by a constant
proportional to the topological winding number associated
with the \lq\lq large'' gauge transformation. To be specific,
consider the three dimensional Chern-Simons action
\begin{equation}
S_{CS} = M \int d^{3}x\,{\rm
Tr}\,\epsilon^{\mu\nu\lambda}\,A_{\mu}(\partial_{\nu}A_{\lambda} +
{2g\over 3} A_{\nu}A_{\lambda})
\end{equation}
where $A_{\mu}$ is a matrix valued gauge field in some representation of
the
gauge algebra. Then,
under  a gauge transformation
$A_{\mu}\rightarrow U^{-1}A_{\mu}U + {1\over g}U^{-1}\partial_{\mu}U$
the Chern-Simons action changes as
\begin{equation}
S_{CS}\rightarrow S_{CS} - {4\pi M\over g^{2}}\;\times 2\pi  W
\end{equation}
where $W$ is the winding number of the gauge transformation defined to
be
\begin{equation}
W = {1\over 24\pi^{2}}\int d^{3}x\,{\rm
Tr}\,\epsilon^{\mu\nu\lambda}\,\partial_{\mu}UU^{-1}\partial_{\nu}UU^{-1}
\partial_{\lambda}UU^{-1}
\end{equation}
The winding number is a topological number (integer) and unless the
gauge transformation belongs to the trivial topology class, it is
clear that the Chern-Simons action will not be gauge
invariant. However, if the coefficient of the
Chern-Simons term is quantized in units of ${g^{2}\over 4\pi}$, then
the path integral, which involves $\exp(i S_{CS})$, will be invariant
and
we can define a consistent quantum theory \cite{DJT}.

Chern-Simons actions can be induced radiatively, with a perturbatively
calculable coefficient. For example, for massive fermions
interacting with a gauge field at zero temperature, radiative
corrections due to the fermions induce a Chern-Simons term with a
coefficient ${1\over 2}$ (in units of ${g^{2}\over 4\pi}$ for every
flavor) \cite{redlich}. Taking into account the intrinsic global anomaly
\cite{W2}, the  effective action is in fact invariant. Alternatively,
we can simply consider an even number of fermion flavors.

These radiative corrections at finite temperature are even more
interesting. At one loop, in the static limit, they induce a
Chern-Simons
term with a temperature dependent coefficient \cite{BDP} such that
\begin{equation}
M\rightarrow M - {g^{2}\over 8\pi}\,{m\over |m|}\,\tanh{\beta|m|\over 2}
\end{equation}
where $\beta$ is the inverse temperature (in units of the Boltzmann
constant). This is now a continuous function of temperature and,
consequently, it can no longer be quantized in units of ${g^{2}\over
4\pi}$ even for an even number of fermion flavors. It would appear,
therefore, that \lq\lq large'' gauge invariance would be lost at
finite temperature which is quite mysterious since gauge invariance
has no direct relation with temperature.

An interesting possible resolution to this puzzle comes from a study of
the Chern-Simons theory in $0+1$ dimensions \cite{DLL} which has all
the  features
of the $2+1$ dimensional theory and yet is much simpler so that the
theory can be exactly solved. It was observed there that, at finite
temperature, the radiative corrections give rise to an infinite number
of non-extensive terms besides the induced Chern-Simons term and
that the effective action due to radiative corrections can be exactly
summed in a closed form. This has to be contrasted with the case at
zero temperature, where the only nontrivial radiative correction was
the Chern-Simons term. Furthermore, it was observed that the summed
effective action at finite temperature is invariant under \lq\lq
large'' gauge transformations once the tree level coefficient is
quantized and we have an even number of fermion flavors. This is,
indeed, quite interesting since it points out that even when the
Chern-Simons term itself may violate \lq\lq large'' gauge invariance,
there may be other terms in the effective action which can compensate to
make the total effective action gauge invariant.

This mechanism extends \cite{DGS,FRS,AF,GF,LLS,HM} to $2+1$ dimensional
Abelian theories  for a restricted class of static backgrounds
$A_\mu=(A_0(t),\vec{A}(\vec{x}))$. However, this is not the full answer,
since for these backgrounds (and for  their trivial non-abelian
generalizations) the \lq\lq large'' gauge  transformations in fact have
zero winding number - the shift in the  Chern-Simons actions comes from
a
total derivative term, not from the winding  number piece. Furthermore,
such backgrounds only address the static limit,  while the non-static
limit is known to be very different
\cite{kao,DD2}.
 Of course, calculations are more
difficult in the $2+1$ dimensional theory and one does not expect to
be able to sum all the terms in the effective action of this
theory. Therefore, to study the problem of \lq\lq large'' gauge
invariance in this theory, we must develop a systematic procedure. The
natural idea, of course, would be to write a Ward identity for \lq\lq
large'' gauge invariance, which  relates different amplitudes
and, therefore, can be perturbatively checked even if the complete
effective action is difficult to evaluate. It is with this in mind
that we have chosen to study the question of the Ward identity for
\lq\lq large'' gauge transformations in the $0+1$ dimensions. Clearly,
the derivation of the Ward identity for gauge transformations which
are topologically nontrivial is hard, but, at least, in $0+1$
dimensions, we have the exact effective actions in closed forms and,
therefore, such theories provide a natural starting point. In this
paper,  we carry
out such a study in detail. In section {\bf 2}, we recapitulate
briefly all the relevant facts known from the studies in the $0+1$
dimensional theories. In section {\bf 3}, we try to derive the
relevant Ward identity for a single flavor fermion theory from the
effective action, directly by brute
force. As we will show, this is quite hard since the Ward identities
are extremely nonlinear. An alternate method is to look at the Ward
identities in terms of the exponential of the effective action which
we do in section {\bf 4}. These identities are linear and easier to
handle. (Of course, the nonlinearity creeps in when we transform back
to the effective action.) The nonlinearity of these identities brings
in many interesting features which we are not used to. Thus, for
example, unlike the Ward identities for small gauge invariance, here
the identities do not obey superposition. Consequently, if a theory
has two distinct sectors, the sum of the effective action coming from
the two sectors does not have the same structure of the identity as
satisfied by the individual
contributions. We point out all such features and present a brief
conclusion in section {\bf 5} pointing out which features are
likely/unlikely to extend to the $2+1$ dimensional theory.

\section{Recapitulation of Results:}

We first recapitulate all the relevant results known from
studies in $0+1$ dimensional theories. Recall that the
theory of massive fermions with $N_{f}$ flavors interacting with an
Abelian  gauge field including a Chern-Simons term is described by the
action (We assume $m>0$ for simplicity.)
\begin{equation}
S_{\rm fermion} = \int dt\,\overline{\psi}(i\partial_{t}-m-A)\psi -
\kappa\int dt\,A\label{1}
\end{equation}
where we have suppressed the flavor index for the fermions and we note
that the last term is the Chern-Simons term in $0+1$ dimension. It is
worth  emphasizing
that even though the gauge field here is Abelian, this theory
has all the properties of a $2+1$ dimensional non-Abelian
theory. Under a gauge transformation,
$\psi\rightarrow e^{-i\lambda}\,\psi$, $ A\rightarrow A +
\partial_{t}\lambda$,
the fermion action is invariant, but the complete action changes:
\begin{equation}
S_{\rm fermion}\rightarrow S_{\rm fermion} - \kappa\,2\pi N
\end{equation}
where $N$ is the appropriate winding number and it is clear that the
tree level coefficient $\kappa$ of the Chern-Simons term must be
quantized for
the theory to be consistent. The mass term for the fermion breaks
charge conjugation invariance and, consequently, the radiative
corrections
due to the
fermions generate a Chern-Simons term at finite temperature modifying
the coefficient as
\begin{equation}
\kappa\rightarrow \kappa - {N_{f}\over 2}\,\tanh {\beta m\over 2}
\end{equation}
This is analogous to the behavior in the $2+1$ dimensional
theory. In particular, we note that, at zero temperature
($\beta\rightarrow\infty$) with an even number of flavors, this is
compatible  with the
quantization of the coefficient of Chern-Simons term, but it poses a
problem at finite temperature suggesting that gauge invariancce may be
violated if the temperature is nonzero.

In this case, of course, the effective action can be exactly evaluated
and has the form \cite{DLL,DD1}
\begin{eqnarray}
\Gamma  & = &  \Gamma_{f}^{(N_{f})} - \kappa a  =   - iN_{f}\log
{\cosh{(\beta m+ia)\over 2}\over \cosh{\beta m\over 2}} - \kappa a\nonumber\\
\noalign{\kern 4pt}%
 & = & - iN_{f}\log\left(\cos
{a\over 2} + i\tanh {\beta m\over 2}\sin {a\over 2}\right) - \kappa
a\label{2}
\end{eqnarray}
where we have defined (We have normalized the effective action so that
it vanishes for $A=0$.)
\begin{equation}
a = \int dt\,A(t)
\end{equation}
There are several things to note from the form of this effective
action. First, this is a non-extensive action (involves powers of an
integrated quantity). Non-extensive actions do not arise at zero
temperature from requirements of locality, but locality is not
necessarily respected at finite temperature. In fact, it is easily seen
from small gauge invariance that, in this theory, if higher order terms
do not vanish, the effective action must be non-extensive. For example,
let us note
\cite{DD1} that
if we have a quadratic term in the effective action, of the form
$S_{q} = {1\over 2}\,\int
dt_{1}\,dt_{2}\,A(t_{1})F(t_{1}-t_{2})A(t_{2})$,
then invariance under small gauge transformations would require
\begin{equation}
\delta S_{q} = \int dt_{1}\,dt_{2}\,\partial_{t_{1}}\lambda
F(t_{1}-t_{2})A(t_{2}) = - \int dt_{1}\,dt_{2}\,\lambda
\partial_{t_{1}}F(t_{1}-t_{2}) A(t_{2}) = 0
\end{equation}
whose general solution is $F=$ constant. If the constant is nonzero,
the quadratic action becomes non-extensive (a quadratic function of
$a$). Thus, one of the important features of this theory is that
$\Gamma_{f}^{(N_{f})}$ and, therefore, $\Gamma$ is a function of $a$.
This
simple feature is unlikely to generalize to the $2+1$ dimensional
theory. In fact, it is already known \cite{DDS1} that this does not
hold even in $1+1$ dimensions because the Ward identities for small
gauge invariance are not restrictive enough.

Another feature to note is that, under a \lq\lq large'' gauge
transformation, for which
\begin{equation}
a\rightarrow a + 2\pi N,
\end{equation}
the effective action coming from the radiative corrections due to the
fermions transforms as
\begin{equation}
\Gamma_{f}^{(N_{f})}(a)\rightarrow \Gamma_{f}^{(N_{f})}(a+2\pi N) =
\Gamma_{f}^{(N_{f})}(a) +  \pi N_{f}N\label{3}
\end{equation}
so that the theory continues to be well defined for an even number of
fermion flavors. That is, even though the coefficient of the
Chern-Simons term is no longer quantized at finite temperature, the
noninvariance of this term is completely compensated for by all the
higher
order terms in the effective action. In fact, the most important thing
to observe in this connection is that the inhomogeneous transformation
of  the fermion effective action is independent of temperature, as we
should expect since gauge transformations are not related to
temperature.

In $0+1$ dimensions, we also know the results for the effective action
of a massive, complex scalar
field interacting with the Abelian gauge field \cite{BD1}. Consider the
theory with action
\begin{equation}
S_{\rm scalar} = \int dt\,
\left((\partial_{t}-iA)\phi^{*}\,(\partial_{t}+iA)\phi -
m^{2}\phi^{*}\phi\right) - \kappa \int dt\,A\label{4}
\end{equation}
where, we have again suppressed the number of flavors for
simplicity. The mass term in this theory does not break parity and,
consequently, there is no Chern-Simons term generated. In fact, at
zero temperature, the radiative corrections due to the scalar fields
identically vanishes which follows from a combination of invariance
under small gauge transformation and the absence of parity
violation. Nonetheless, at finite temperature, the effective action
coming from the scalar fields is nontrivial and has the form
\begin{eqnarray}
\Gamma_{s}^{(N_{f})} & = & iN_{f}\log{\sinh{(\beta m+ia)\over
2}\sinh{(\beta m-ia)\over 2}\over \sinh^{2}{\beta m\over
2}} =  i N_{f} \log\left(\cos^{2} {a\over 2} + \coth^{2}
{\beta m\over 2}\,\sin^{2} {a\over 2}\right)\nonumber\\
\noalign{\kern 4pt}%
  & = & i N_{f} \log\left({(\cosh \beta m - \cos a)\over
2\sinh^{2}(\beta m/2)}\right)\label{5}
\end{eqnarray}
We see that even though there is no Chern-Simons term, we
would have run into the problem of \lq\lq large'' gauge invariance had
we done a perturbative calculation and looked at the quadratic terms
alone. The effective action, once again, is  non-extensive and is a
function of $a$, namely, $\Gamma_{s}=\Gamma_{s}(a)$. Furthermore,
under a large gauge transformation,
\begin{equation}
\Gamma_{s}^{(N_{f})}(a)\rightarrow \Gamma_{s}^{(N_{f})}(a+2\pi N) =
\Gamma_{s}^{(N_{f})}(a)\label{5'}
\end{equation}
Namely, the action is invariant independent of the temperature.

Finally, consider a simple supersymmetric model in $0+1$
dimensions, with action \cite{BD1}
\begin{eqnarray}
S_{\rm super} & = & \int
dt\,\left((\partial_{t}-iA)\phi^{*}\,(\partial_{t}+iA)\phi -
m^{2}\phi^{*}\phi +
\overline{\psi}(i\partial_{t}-m-A)\psi\right)\nonumber\\
\noalign{\kern 4pt}%
 &  & + \int dt \left({1\over 2}(A+\dot{\theta})^{2} + {i\over
2}(\lambda+\xi)(\dot{\lambda}+\dot{\xi})\right) - \kappa \int
dt\,A\label{6}
\end{eqnarray}
Here, in addition to the usual scalar and fermionic fields with
identical number of flavors, we also have a stuckelberg multiplet of
fields. The effective action for this theory is quite simple
\begin{equation}
\Gamma = \Gamma_{susy}^{(N_{f})}(a) + \int dt\left({1\over
2}(A+\dot{\theta})^{2}+{i\over
2}(\lambda+\xi)(\dot{\lambda}+\dot{\xi})\right) - \kappa \int dt\,A
\end{equation}
where we recognize that
\begin{equation}
\Gamma_{susy}^{(N_{f})}(a) =  \Gamma_{f}^{(N_{f})}(a) +
\Gamma_{s}^{(N_{f})}(a)
 = - iN_{f}\log{2\sinh^{2}{\beta m\over 2}(\cos{a\over 2} +
i\tanh{\beta m\over 2}\sin{a\over 2})\over (\cosh\beta m - \cos
a)}\label{7}
\end{equation}
and the transformation properties follow from our earlier
discussion. With these basics, we are now ready to get into the
question of the Ward identity for \lq\lq large'' gauge invariance.

\section{Ward Identity (Hard Way):}

To begin with, let us consider the model for a single (flavor) massive
fermion interacting with an Abelian gauge field. The Lagrangian is
trivially obtained from eq. (\ref{1}). Let us denote by
$\Gamma_{f}^{(1)}$ the effective action which results from
integrating out the fermions. From the general arguments of the last
section, we know that this effective action will be a function of $a$
such that under a large gauge transformation
\begin{equation}
\Gamma_{f}^{(1)} (a)\rightarrow \Gamma_{f}^{(1)}(a+2\pi N) =
\Gamma_{f}^{(1)}(a) + \pi N
\end{equation}
The Taylor expansion of this relation gives
\begin{equation}
\sum_{n=1}^{\infty}\sum_{m=0}^{\infty} {a^{m}(2\pi N)^{n}\over
m!n!}\,\left.{\partial^{n+m}\Gamma_{f}^{(1)}\over \partial
a^{n+m}}\right|_{a=0} = \pi N
\end{equation}

This is an infinite number of constraints which can also be rewritten
in the form
\begin{eqnarray}
\sum_{n=1}^{\infty} {(2\pi N)^{n}\over
n!}\,\left.{\partial^{n}\Gamma_{f}^{(1)}\over \partial
a^{n}}\right|_{a=0} & = & \pi N\nonumber\\
\noalign{\kern 4pt}%
{a^{m}\over m!}\,\sum_{n=1}^{\infty} {(2\pi N)^{n}\over
n!}\,\left.{\partial^{n+m}\Gamma_{f}^{(1)}\over \partial
a^{n+m}}\right|_{a=0} & = & 0\quad {\rm for}\quad m>0\label{7'}
\end{eqnarray}
Solving this set of equations is tedious, but with a little bit of
work, we can determine that the simplest relation which will satisfy
the infinity of relations has the form (This is, however, not the most
general relation as we will discuss later in the case of the
supersymmetric
theory.)
\begin{equation}
{\partial^{2}\Gamma_{f}^{(1)}\over \partial a^{2}} = i\left({1\over 4}
- \left({\partial\Gamma_{f}^{(1)}\over \partial
a}\right)^{2}\right)\label{8}
\end{equation}
This relation (as well as the ones following from it) can be checked
explicitly for the first few low order amplitudes of the theory
\cite{DD1,BD2} and,
consequently, eq. (\ref{8}) can be thought of as the Ward identity for
\lq\lq
large'' gauge invariance (or better yet, the Master equation from
which all the relevant relations can be obtained by taking further
derivatives with respect to $a$). This relates higher point functions
to lower ones as we would expect a Ward identity to do and must hold at
zero as well as nonzero temperature. However, unlike conventional Ward
identities associated with small gauge invariance, we note that this
relation  is nonlinear. In some sense, this is to be expected for
\lq\lq large'' gauge transformations and we comment on the
consequences of this feature later.

We  note from the explicit form of the effective action
determined earlier (see eq. (\ref{2}) for $N_{f}=1$) that
eq. (\ref{8})  indeed holds for the single flavor fermion theory. Let
us  also note here that all the higher point functions can
be determined from the Ward identity in eq. (\ref{8}) in terms of the
one point function which has to be calculated from the theory and is
known to be \cite{DD1,BD2}
\begin{equation}
\left.{\partial\Gamma_{f}^{(1)}\over \partial a}\right|_{a=0} =
{1\over 2}\,\tanh {\beta m\over 2}
\end{equation}
It follows from this (using the identity in eq. (\ref{8})) that, at
zero temperature, the two point and all other higher point functions
vanish.  In
fact, from the Ward identity (\ref{8}), we can determine the form of
the effective action to be (recall that $\Gamma_{f}(a=0)=0$)
\begin{equation}
\Gamma_{f}^{(1)}(a) = -i \log\left(\cos{a\over 2} +
2i\left.{\partial\Gamma_{f}^{(1)}\over \partial
a}\right|_{a=0}\,\sin{a\over 2}\right)
\end{equation}
Namely, the effective action can be completely determined from the
knowledge of the one point function which, of course, coincides with
the exact calculations. It is clear, however, that this way of
deriving the Ward identity is extremely hard and, in particular, if
the theory is complicated (remember that so far we have only
considered a single flavor of massive fermion), then, it may be
much more difficult to determine the Ward identity.

\section{Simple Derivation of the Ward Identity:}

To find a simpler way of deriving the \lq\lq large'' gauge Ward
identity, let us define
\begin{equation}
\Gamma(a) = \mp i \log W(a)\label{9}
\end{equation}
where the upper sign is for a fermion theory while the lower one
corresponds to a scalar theory. Namely, we are interested in looking at
the
exponential of the effective action ({\it i.e.} up to a factor of $i$,
$W$ is the basic determinant that would arise from integrating out a
particular
field). Once again, we will restrict ourselves to a single flavor
massive fermion or a single flavor massive complex scalar since any
other theory can be obtained from these basic components. The
advantage of studying $W(a)$ as opposed to the effective action lies
in the fact that, in order for $\Gamma(a)$ to have the right
transformation
properties  under a large gauge transformation (see
eqs. (\ref{3}),(\ref{5})), $W(a)$ simply has to be
quasi-periodic. Consequently, from the study of harmonic oscillator (as
well as Floquet theory), we see that $W(a)$ has to satisfy a simple
equation of the form
\begin{equation}
{\partial^{2}W(a)\over \partial a^{2}} + \nu^{2}\,W(a) = g\label{10}
\end{equation}
where $\nu$ and $g$ are parameters to be determined from the
theory. In particular, let us note that the constant $g$ can depend on
parameters of the theory such as temperature whereas we expect the
parameter
$\nu$, also known as the characteristic exponent, to be independent of
temperature and equal to an odd half integer for a fermionic mode or
an integer for a scalar mode. However, all these properties should
automatically result from the structure of the theory. Let us also
note here that the relation (\ref{10}) is simply the equation for a
forced oscillator whose solution has the general form
\begin{equation}
W(a) = \frac{g}{\nu^2} + A\cos(\nu a + \delta) = \frac{g}{\nu^2} +
\alpha_{1}\cos\nu a +
\alpha_{2}\sin\nu a\label{10''}
\end{equation}
The constants $\alpha_{1}$ and $\alpha_{2}$ appearing in the solution
can again be determined from the theory. Namely, from the relation
between $W(a)$ and $\Gamma(a)$, we recognize that we can identify
\begin{eqnarray}
\nu^{2}\alpha_{1} & = & - \left.{\partial^{2}W\over \partial
a^{2}}\right|_{a=0} = \left.\left(\left({\partial\Gamma\over \partial
a}\right)^{2} \mp i\,{\partial^{2}\Gamma\over \partial
a^{2}}\right)\right|_{a=0}\nonumber\\
\noalign{\kern 4pt}%
\nu\alpha_{2} & = & \left.{\partial W\over \partial a}\right|_{a=0} =
\pm i\,\left.{\partial\Gamma\over \partial a}\right|_{a=0}\label{10'}
\end{eqnarray}
From the general properties of the scalar and fermion theories we have
discussed, we intuitively expect $g_{f}=0$ and
$\alpha_{2,s}=0$. However, these should really follow from the
structure of the theory and they do, as we will show shortly.

The identity (\ref{10}) is a linear relation as opposed to the Ward
identity (\ref{8}) in terms of the effective action, and holds both for
a fermionic as well as a scalar mode. In fact, rewriting this in terms
of the effective action (using eq. (\ref{9})), we have
\begin{equation}
{\partial^{2}\Gamma(a)\over \partial a^{2}} =
\pm i\left(\nu^{2}-\left({\partial\Gamma(a)\over \partial
a}\right)^{2}\right) \mp i g\,e^{\mp i\Gamma(a)}\label{11}
\end{equation}
This is reminiscent of the identity in eq. (\ref{8}), but is not
identical. So, let us investigate this a little
bit  more in detail, first for a fermionic mode. In this case, we know
that the fermion mass term breaks parity and, consequently, the
radiative corrections would generate a Chern-Simons term, namely, in
this theory, we expect the one-point function to be
nonzero. Consequently, by taking  derivative of eq. (\ref{11}) (as
well as remembering that $\Gamma(a=0)=0$), we determine
\begin{eqnarray}
(\nu_{f}^{(1)})^{2} & = & \left[\left({\partial\Gamma_{f}^{(1)}\over
\partial
a}\right)^{2} - 3i\,{\partial^{2}\Gamma_{f}^{(1)}\over \partial a^{2}}
- \left({\partial\Gamma_{f}^{(1)}\over \partial
a}\right)^{-1}\left({\partial^{3}\Gamma_{f}^{(1)}\over \partial
a^{3}}\right)\right]_{a=0}\nonumber\\
\noalign{\kern 4pt}%
g_{f}^{(1)} & = & -\left[2i\,{\partial^{2}\Gamma_{f}^{(1)}\over \partial
a^{2}} + \left({\partial\Gamma_{f}^{(1)}\over \partial
a}\right)^{-1}\left({\partial^{3}\Gamma_{f}^{(1)}\over \partial
a^{3}}\right)\right]_{a=0}\label{12}
\end{eqnarray}
This is quite interesting, for it says that the two parameters in
eq. (\ref{10}) or (\ref{11}) can be determined from a perturbative
calculation. Let us note here some of the perturbative results in this
theory \cite{DD1,BD2}, namely,
\begin{eqnarray}
\left.{\partial\Gamma_{f}^{(1)}\over \partial a}\right|_{a=0} & = &
{1\over 2}\,\tanh {\beta m\over 2}\nonumber\\
\noalign{\kern 4pt}%
\left.{\partial^{2}\Gamma_{f}^{(1)}\over \partial a^{2}}\right|_{a=0}
& = & {i\over 4}\,{\rm sech}^{2} {\beta m\over 2}\nonumber\\
\noalign{\kern 4pt}%
\left.{\partial^{3}\Gamma_{f}^{(1)}\over \partial a^{3}}\right|_{a=0}
& = & {1\over 4}\,\tanh {\beta m\over 2}\,{\rm sech}^{2} {\beta m\over
2}
\end{eqnarray}
Using these, we immediately determine from eq. (\ref{12}) that
\begin{equation}
(\nu_{f}^{(1)})^{2} = {1\over 4},\quad\quad g_{f}^{(1)} = 0\label{13}
\end{equation}
so that the equation (\ref{11}) coincides with (\ref{8}) for a single
fermion flavor. Furthermore, we determine now from eq. (\ref{10'})
\begin{equation}
\alpha_{1,f}^{(1)} = 1,\quad\quad \alpha_{2,f}^{(1)} = \pm i\,\tanh
{\beta m \over 2}
\end{equation}
The two signs in of $\alpha_{2,f}^{(1)}$ simply corresponds to the two
possible signs of $\nu_{f}^{(1)}$. With this then, we can solve for
$W(a)$ in  the single flavor fermion  theory and we have (independent
of the sign of $\nu_{f}^{(1)}$)
\begin{equation}
W_{f}^{(1)}(a) = \cos{a\over 2} + i\,\tanh{\beta m\over 2}\sin{a\over
2}\label{13'}
\end{equation}
which can be compared with eq. (\ref{2}).

For a scalar theory, however, we know that the mass term does not
break parity. Consequently, we do not expect a Chern-Simons term to be
generated simply from symmetry arguments. In fact, in the scalar
theory with parity as a symmetry, there cannot be any odd terms (in
$a$) in  the effective action. Consequently, taking derivatives of
eq. (\ref{11}) and keeping this in mind, we obtain
\begin{eqnarray}
(\nu_{s}^{(1)})^{2} & = &
\left[3i\,{\partial^{2}\Gamma_{s}^{(1)}\over \partial a^{2}} -
\left({\partial^{2}\Gamma_{s}^{(1)}\over \partial
a^{2}}\right)^{-1}\,{\partial^{4}\Gamma_{s}^{(1)}\over \partial
a^{4}}\right]_{a=0}\nonumber\\
\noalign{\kern 4pt}%
g_{s}^{(1)} & = & \left[2i\,{\partial^2\Gamma_{s}^{(1)}\over \partial
a^{2}} - \left({\partial^{2}\Gamma_{s}^{(1)}\over \partial
a^{2}}\right)^{-1}\,{\partial^{4}\Gamma_{s}^{(1)}\over \partial
a^{4}}\right]_{a=0}\label{14}
\end{eqnarray}
Once again, we see that the two parameters in the Ward identity can be
determined from the first two nontrivial amplitudes of the theory. For
the scalar theory, the necessary nontrivial amplitudes can be easily
computed \cite{BD1}. (Calculationally, the
scalar theory coincides with two fermionic theories with masses of
opposite sign if we neglect the negative sign associated with fermion
loops. The amplitudes can also be determined from the explicit form of
the effective action in eq. (\ref{5}).)
\begin{eqnarray}
\left.{\partial^{2}\Gamma_{s}^{(1)}\over \partial a^{2}}\right|_{a=0}
& = & {i\over 2\sinh^{2}(\beta m/2)}\nonumber\\
\noalign{\kern 4pt}%
\left.{\partial^{4}\Gamma_{s}^{(1)}\over \partial a^{4}}\right|_{a=0}
& = & - {i\over 2\sinh^{2}(\beta m/2)}\left(1+{3\over 2\sinh^{2}(\beta
m/2)}\right)\label{15}
\end{eqnarray}
It follows from this that
\begin{equation}
(\nu_{s}^{(1)})^{2} = 1,\quad\quad 
g_{s}^{(1)} = {\cosh \beta m\over 2\sinh^{2}(\beta m/2)}\label{16}
\end{equation}
This is indeed consistent with our expectations. Furthermore, we now
determine from eq. (\ref{10'})
\begin{equation}
\alpha_{1,s}^{(1)} = - {1\over 2\sinh^{2}(\beta m/2)},\quad\quad 
\alpha_{2,s}^{(1)} = 0
\end{equation}
so that we can write
\begin{equation}
W_{s}^{(1)}(a) = {(\cosh\beta m - \cos a)\over 2\sinh^{2}(\beta m/2)}
\end{equation}
This can be compared with eq. (\ref{5}).

Thus, we see that the Ward identities for the single flavor fermion
and scalar theories are given, in terms of the effective actions,
respectively by
\begin{eqnarray}
{\partial^{2}\Gamma_{f}^{(1)}\over \partial a^{2}} & = &
i\left({1\over 4}-\left({\partial\Gamma_{f}^{(1)}\over \partial
a}\right)^{2}\right)\nonumber\\
\noalign{\kern 4pt}%
{\partial^{2}\Gamma_{s}^{(1)}\over \partial a^{2}} & = & -
i\left(1-\left({\partial\Gamma_{s}^{(1)}\over \partial
a}\right)^{2}\right) +  {i\cosh\beta m\over 2\sinh^{2}(\beta
m/2)}\,e^{i\Gamma_{s}^{(1)}}
\end{eqnarray}
As we have already pointed out, these are nonlinear identities and,
therefore, superposition does not hold. In fact, even if we are
considering only fermions (or scalars) of $N_{f}$ flavors, the
identity modifies in a nontrivial manner (which can be derived by
simply noting that $W^{(1)}=e^{\pm(i/N_{f})\Gamma^{(N_{f})}}$), namely,
\begin{eqnarray}
{\partial^{2}\Gamma_{f}^{(N_{f})}\over \partial a^{2}} & = &
iN_{f}\left({1\over 4}-{1\over
N_{f}^{2}}\left({\partial\Gamma_{f}^{(N_{f})}\over \partial
a}\right)^{2}\right)\nonumber\\
\noalign{\kern 4pt}%
{\partial^{2}\Gamma_{s}^{(N_{f})}\over \partial a^{2}} & = &
-iN_{f}\left(1-{1\over
N_{f}^{2}}\left({\partial\Gamma_{s}^{(N_{f})}\over \partial
a}\right)^{2}\right) + {iN_{f}\cosh\beta m\over 2\sinh^{2}(\beta
m/2)}\,e^{{i\over N_{f}}\,\Gamma_{s}^{(N_{f})}}
\end{eqnarray}
Incidentally, let us note that although the Ward identity is linear in
$W(a)$, for $N_{f}$ flavors $W^{(N_{f})}$ is a {\it product} of
$W^{(1)}$'s, not a sum, and so superposition is lost.

It is clear, therefore, that while the Ward identity is simple for the
basic fermion and scalar modes, for arbitrary combinations of these
modes, the identity is bound to be much more complicated. However, we
can
still derive the Ward identity from the basic identities for a single
flavor fermion and scalar theory. Thus, as a simple example, let us
consider the supersymmetric theory discussed in section {\bf 2} (see
eq. (\ref{6}-\ref{7})) for a single flavor. In this case, we have
simply a sum of a fermionic and a complex bosonic degree of freedom,
and  defining
\begin{equation}
\Gamma_{susy}^{(1)}(a) = - i \log W_{susy}^{(1)}(a)
\end{equation}
where
\begin{equation}
W_{susy}^{(1)}(a) = {W_{f}^{(1)}(a)\over W_{s}^{(1)}(a)}
\end{equation}
we see that we can no longer write a single identity for
$W_{susy}^{(1)}(a)$. Rather, we will have a coupled set of identities,
one of which, say the one for the fermions will be decoupled.


On the other hand, since the fermion equation is uncoupled and can be
solved (see eq. (\ref{13'})), we can use the solution to write a single
identity for $W_{susy}^{(1)}(a)$ and, therefore, $\Gamma_{susy}^{(1)}(a)$:
\begin{eqnarray}
i\,{\partial^{2}\Gamma_{susy}^{(1)}\over \partial a^{2}} +
\left({\partial\Gamma_{susy}^{(1)}\over \partial a}\right)^{2} -
{3\over 4} & = & \tanh(\frac{\beta m+ia}{2})\,{\partial
\Gamma_{susy}^{(1)}\over
\partial a}\nonumber\\
\noalign{\kern 4pt}%
 &  & - {\cosh\beta m\cosh{\beta m\over 2}\over 2\sinh^{2}{\beta
m\over 2}\cosh{(\beta m+ia)\over 2}}\,e^{i\Gamma_{susy}^{(1)}}\label{18}
\end{eqnarray}
This is very different from eq. (\ref{8}) and yet satisfies the
infinite set of constraints in eq. (\ref{7'}). (Let us note here that
the invariance properties of $\Gamma_{susy}^{(1)}(a)$ is the same as
that for a single flavor fermion theory.) Thus, as we had mentioned
earlier, the identity for a basic single flavor theory is simpler. The
identities following from eq. (\ref{18}) can be perturbatively checked.
In
fact, the identity can even be solved in the following way. Consider
eq. (\ref{18}). Using the method of Fourier decomposition, it is seen
after some algebra
that the solution to eq. (\ref{18}) has the form
\begin{equation}
W_{susy}^{(1)}(a) = (1-e^{-\beta m})\sum_{k=0}^{\infty} e^{-k\beta
m}\left[\cos(k+{1\over 2})a + i\tanh^{2}{\beta m\over 2}\sin(k+{1\over
2})a\right]\label{19}
\end{equation}
It is interesting that even for this simple model, which contains just
a single flavor of fermion and a complex scalar, $W(a)$ becomes a sum
over infinitely many distinct Fourier modes as opposed to the case of
either a single fermion or a single complex scalar where $W(a)$
involves only a single Fourier component. Let us note here that,
although $W_{susy}^{(1)}(a)$ in eq. (\ref{19}) appears different from
that in eq. (\ref{7}), the series in (\ref{19}) can, in fact, be
summed \cite{GR} and
coincides with the result in (\ref{7}) for a single flavor.

\section{Conclusion:}

In this paper, we have systematically studied the question of \lq\lq
large''
gauge invariance, in  the $0+1$ dimensional Chern-Simons theory.
The effective actions, in $0+1$ dimensions, are
functions of $a=\int dt A(t)$ which is a consequence of Ward
identities for small gauge invariance and makes the derivation of the
\lq\lq large'' gauge identities relatively simple. This is a feature
that we do not expect to generalize to $2+1$ dimensions, except for
special
static backgrounds. Explicitly,
we have derived the \lq\lq large'' gauge Ward
identities for a single flavor fermion theory as well as a single
flavor complex scalar theory interacting with an Abelian gauge
field. These identities are simple, but nonlinear. The Ward identity
for any other theory can be derived from them and is, in general, more
complicated because of the nonlinear nature of the identities. In
particular, we have shown that the solutions of the Ward identity for a
single flavor fermion theory or a single flavor complex scalar theory
involves a single characteristic index and is simple while for a more
complex theory (even for a sum of just a single fermion and a single
scalar),  it  involves a sum over an infinity of Fourier
modes. This is a feature which we believe would generalize to $2+1$
dimensions.

\section*{Acknowledgments}

This work was supported in part by the U.S. Dept. of Energy Grant
DE-FG 02-91ER40685, NSF-INT-9602559 as well as CNPq. GD acknowledges the
support of the DOE under grant DE-FG02-92ER40716.00, and thanks the
Technion Physics Department for its hospitality.

\end{document}